\documentclass[11pt,a4paper]{article}
\usepackage[margin=2.5cm]{geometry}

\usepackage[T1]{fontenc}
\usepackage{lmodern}
\usepackage[utf8]{inputenc}

\usepackage{amsmath, amsfonts, amssymb, amsthm}
\usepackage{mathtools}
\usepackage{bm}
\usepackage{nicefrac, xfrac}
\usepackage[scr=rsfso]{mathalfa}

\usepackage{siunitx}
\providecommand{\angstrom}{\text{Å}}
\DeclareSIUnit\angstrom{\angstrom}

\usepackage{graphicx}
\usepackage{booktabs}
\usepackage{array}
\usepackage{multirow}
\usepackage{multicol}
\usepackage{tabularx}
\usepackage{float}
\usepackage{subcaption}
\usepackage{enumitem}
\usepackage{soul}
\graphicspath{{figures/}}

\newcommand{\x}{{\bf x}}
\newcommand{\R}{\mathbb{R}}
\newcommand{\M}{\mathbf{M}}
\newcommand{\I}{\mathbf{I}}
\newcommand{\J}{\mathbf{J}}

\usepackage{authblk}
\usepackage{hyperref}

\newtheorem{lemma}{Lemma}
\theoremstyle{remark}
\newtheorem{remark}{Remark}
\newtheorem{definition}{Definition}

\title{Particle picking with an earth mover's seminorm}

\author[a]{Dvir  Moshe}
\author[b]{Amit Moscovich}
\author[c]{Amit Singer}
\author[a]{Ayelet Heimowitz}

\affil[a]{Department of Electrical and Electronics Engineering, Ariel University, Ariel, Israel}
\affil[b]{Department of Statistics and Operations Research, Tel Aviv University, Tel Aviv, Israel}
\affil[c]{Department of Mathematics, Program in  Applied and Computational Mathematics, Princeton University, Princeton, USA}

\date{} 

\begin{document}
\maketitle

\begin{abstract}
Particle picking is a key step in single-particle cryo-electron microscopy.
It is a challenging task where tomographic projections of proteins and other macromolecules are identified in extremely noisy images. In this work, we introduce a particle picker based on a seminorm defined in terms of a fast wavelet approximation to the earth mover's distance.
Our picker is template-free, self-supervised and does not require any prior knowledge of the molecules of interest other than their approximate size.
We test it on three public datasets and show that it achieves state-of-the-art reconstructions.
\end{abstract}

\section{Introduction}

\begin{figure*}[h]
    \centering
    \begin{subfigure}{0.4\textwidth}
        \includegraphics[width=\linewidth]{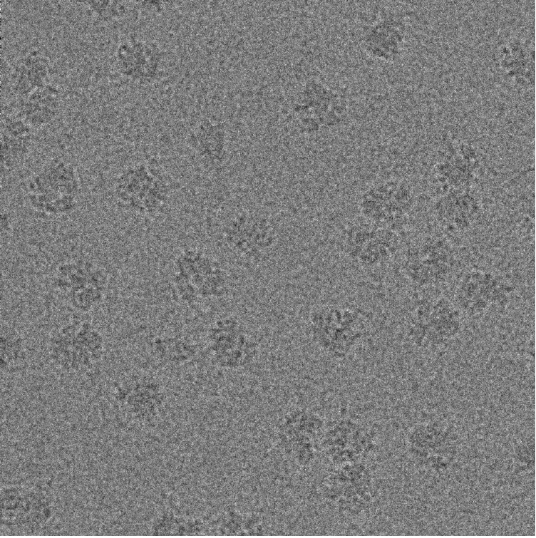}
        \caption{}
        \label{subfig:micrograph}
    \end{subfigure}
    \hspace{1cm}
    \begin{subfigure}{0.4\textwidth}
        \includegraphics[width=\linewidth]{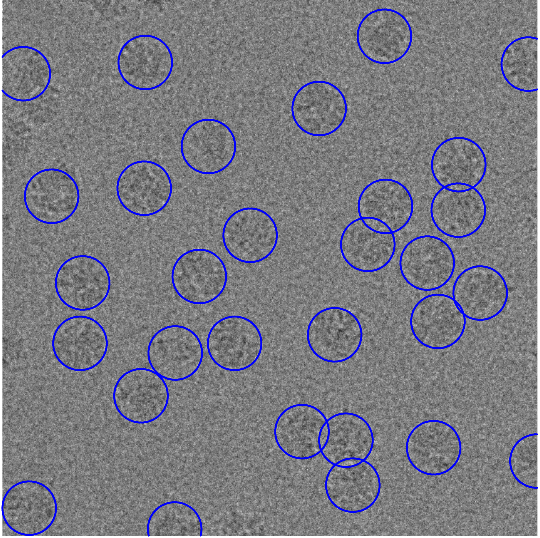}
        \caption{}
        \label{subfig:pick_example}
    \end{subfigure}
    \caption{Results of our suggested framework on a micrograph of the 80S ribosome. (a) Part of a micrograph from the EMPIAR-10028 dataset; (b)  Same image with picked particles.}
    \label{fig:emd_results}
\end{figure*}

Single-particle cryo-electron microscopy (cryo-EM) is a widely used technique for reconstructing 3-D density maps of macromolecules at near-atomic resolution~\cite{bai2015cryo, deoliveira2021}. In this method, frozen-hydrated specimens are imaged using an electron microscope that shoots a high-energy electron beam through the ice and the macromolecules embedded within, creating 2-D projections of the macromolecules of interest, taken from random orientations. These 2-D projections are captured in an image known as a micrograph.
See Figure~\ref{subfig:micrograph} for an example of a micrograph taken from EMPIAR-10028 with a sample of 80S ribosomes \cite{iudin2023empiar, wong2014cryo}.
To reconstruct high resolution 3-D densities, tens of thousands to millions of these projections must be identified and extracted during the \emph{particle picking} step of the cryo-EM computational pipeline~\cite{Frank2006, sigworth2004,grigorieff2011, brilot2012, scheres2015}. 

One of the main difficulties in particle picking is the high level of noise present in the micrographs. During the imaging process, the electrons that pass through the sample may experience elastic or inelastic scattering. The inelastic scattering events cause radiation damage to the specimen. Therefore, to preserve high resolution features, the electron dose must be severely limited~\cite{henderson1992, baker2010cryo, brilot2012, BendoryBartesaghiSinger2019,singer2020}. It is this limitation that causes the notoriously low signal-to-noise ratio (SNR) of the micrographs and, thereby, makes the task of particle picking challenging~\cite{sigworth2004, scheres2015}.

Approaches to particle picking can be divided into three main categories. The first category consists of methods based on template matching. The templates may be obtained by computing 2-D projections of previously published 3-D reconstructions of similar macromolecules, mathematical functions such as Gaussians, etc.~\cite{frank1983, Frank2006, sigworth2004, nicholson2009, vos2009, zhang2017}. These methods are typically prone to high levels of false detections due to the low SNR in the micrograph~\cite{sigworth2004, singer2020}. The low SNR in the micrographs also  leads to localization issues. That is, picked particles are not centered. Lastly,  template-based methods are susceptible to the \emph{Einstein from noise} effect~\cite{shatsky2009, henderson2013avoiding} where an erroneous template, such as an image of Albert Einstein, is recreated in the 2-D class averages, even when the micrographs are pure noise.  The second line of work is based on neural networks. These networks are often laborious to train as they require human-annotated micrographs. Additionally, the difference in human pickers may result in varying qualities of  reconstructions~\cite{wang2016deeppicker,nguyen2018,bepler2019positive,wagner2019}. 
The third category of particle pickers assumes a specific noise model, such as independent and identically distributed additive Gaussian noise
\cite{sigworth2004}. In this manner, no assumptions are made regarding the projections, only the noise and artifacts~\cite{heimowitz2018, eldar2020}.

In this paper we present a method for self-supervised particle picking that falls under the third category. It identifies areas of pure noise in the micrograph using a seminorm based on the earth mover's distance (EMD), also known as the 1-Wasserstein metric. This seminorm  tends to assign lower values to areas of pure noise as compared to areas that contain projections of the macromolecule of interest, allowing us to automatically select a training set for a binary classifier which we then use for segmenting the micrograph into noise and projection pixels. 

Our approach is tested on both synthetic data and real-world datasets taken from EMPIAR~\cite{iudin2023empiar}. We show that, while our method is simple, template-free and self-supervised, its performance is nonetheless comparable to state-of-the-art particle pickers. See Figure~\ref{subfig:pick_example} for an example of particles picked by our method on an 80S ribosome dataset \cite{wong2014cryo}.

\section{Background} \label{sec:backgrounds}

\subsection{Image Formation Model}
\label{subsec:ifm}

We denote the Coulomb potential of the 3-D macromolecule as $v: \mathbb{R}^3 \rightarrow \mathbb{R}$. A clean 2-D projection is denoted by  $Pv$ and defined as follows~\cite{singer2020, kapulin2025}:
\begin{equation}
 (P v)(x,y)  = \int_{-\infty}^{\infty} v \left(x,y,z \right) dz.
\end{equation}
Each molecule is frozen at a random orientation.
We denote by $R_i$ its $3 \times 3$ rotation matrix such that $(R_i \circ v)({\bf r}) = v(R_i^{-1} {\bf r})$.
The clean projection is therefore $P R_i v$.
Under the weak phase approximation \cite{Frank2006, BhamreZhangSinger2016, singer2020, kapulin2025}, the $i$-th 2-D projection in a micrograph is modeled as
\begin{equation}
    \I_i = c_i h_i * T_{t_i} P R_i v + \sigma_i \eta_i,
    \label{equ:image_formation}
\end{equation}
where $h_i$ is a point spread function that is convolved with the clean projection, the scalar term $c_i$ is a contrast amplitude \cite{ShiSinger2022},
$T_{t_i}$ is a 2-D translation by $t_i \in \mathbb{R}^2$, and
$\eta_i$ accounts for shot noise and is typically modeled as colored Gaussian noise. The noise variance, determined by~$\sigma_i$, changes across the micrograph according to variations in the thickness of the ice and so does $c_i$.
For simplicity we assume that the point spread function is fixed throughout the micrograph.\footnote{In practice changes in ice thickness will cause changes in the point spread function. However, per particle CTF estimation is  usually dealt with downstream of the reconstruction.}

\subsection{Wasserstein metrics on images}
\label{subsec:emd}

Consider an $n \times n$ image $\I \in \mathbb{R}_{\ge 0}^{n \times n}$, with pixel coordinates on the regular grid $\{0, 1, \ldots, n-1\}^2$.
We identify images with discrete measures, where each pixel intensity gives the mass at the corresponding grid point.

\begin{definition}[admissible coupling]
A \emph{coupling} between a source image~$\I$ and a target image~$\J$ is a non-negative array $\pi_{ij}^{k\ell}$ that prescribes the amount of mass transported from each source pixel $(i,j)$ to every target pixel $(k, \ell)$.
An \emph{admissible coupling} (or transport plan) is a coupling that accounts for the entire mass of the source and target, i.e.
\begin{align}
    \forall i,j: \ I_{ij} = \sum_{k, \ell} \pi_{ij}^{k\ell}
    \qquad \text{and} \qquad
    \forall k,\ell: \ J_{k\ell} = \sum_{i,j} \pi_{ij}^{k\ell}.
\end{align}
\end{definition}
We denote the set of all admissible couplings between $\I$ and $\J$ by $\Pi({\I}, {\J})$.
For a coupling to exist the images must have the same total mass, so it is typically assumed that all images have been normalized to have unit mass.

If one associates a ground cost $C_{ij}^{k\ell}$ with the transportation of a unit of mass from $(i,j)$ to $(k,\ell)$, then under a linear model the  cost of a transport plan $\pi$ is $\langle C, \pi \rangle = \sum_{ijk\ell} C_{ij}^{k\ell} \pi_{ij}^{k\ell}$.
The Wasserstein metric is defined as the minimal cost of an admissible coupling. 
\begin{definition}[Wasserstein metric]
    Let $d_{ij}^{k\ell}$ be a distance metric on the pixel grid and let $p \ge 1$ be a constant.
    The Wasserstein-$p$ metric is defined as the cost of a minimizing transport plan with respect to the ground cost $C = d^p$,
    \begin{align}
        W_p({\bf I}, {\bf J}) := \left( \min_{\pi \in \Pi({\bf I}, {\bf J}) } \langle d^p, \pi \rangle \right)^{1/p}.
    \end{align}
\end{definition}
The $p=1$ case is also known as the \emph{earth mover's distance} (EMD).
It is the least amount of work necessary to transport the mass distribution of  image ${\bf I}$ so that it matches that of  image ${\bf J}$, where work is defined as mass times distance moved.
On $n \times n$ images, the EMD is typically computed by solving a linear program with $n^4$ variables, one for each pair of source and target pixels. As this is rather time-consuming it is common to
turn to approximations such as the wavelet EMD~\cite{shirdhonkar2008,CraigYu2025},
entropy-regularized variants \cite{Cuturi2013}, or the sliced Wasserstein distance~\cite{Bonneel2014, Kolouri2019, ShiSingerVerbeke2026}. 

\subsection{Wavelet earth mover's distance and seminorm} \label{sec:WEMD}

Let $\mathcal{W}$ be a discrete wavelet transform (DWT) of an image in some orthonormal wavelet basis and let $L$ be the wavelet decomposition level.
We index the wavelet coefficients by tuples $(\ell, k, s_x, s_y)$ where $\ell=0, \ldots, L-1$ is the scale (with larger values corresponding to finer details), $k \in \{1,2,3\}$ indicates whether the wavelet is  horizontal, vertical, or mixed; and $s_x, s_y$ are the $x$ and $y$
shifts.

The \emph{wavelet earth mover's distance} (WEMD) between a pair of 2-D images $ \I, \J  \in \R^{n \times n}$ is a weighted $\ell_1$ distance of wavelet detail coefficients, 
\begin{align} \label{eq:WEMD_definition}
    \mathrm{WEMD} \left(\I, \J\right)
    :&= 
    \sum_{\ell, k, s_x, s_y}
    2^{-2\ell}
    |
        (\mathcal{W} \I)[\ell, k, s_x, s_y]
        -
        (\mathcal{W} \J)[\ell, k, s_x, s_y]
    | \nonumber
    \\
    &=
    \sum_{\ell, k, s_x, s_y}
    2^{-2\ell}
    |
        (\mathcal{W} (\I - \J))[\ell, k, s_x, s_y]
    |,
\end{align}
where the sum is over all detail coefficients of all scales $\ell=0, \ldots, L-1$.
The WEMD is a surrogate for the Wasserstein-1 metric between images \cite{shirdhonkar2008,CraigYu2025} that is extremely fast since the DWT is computable in linear time.
Based on this, we define a wavelet-based seminorm for (signed) images: 

    \begin{definition}[WEMD seminorm] \label{def:wemd_seminorm}
    Let $\I \in \R^{n \times n}$ be an $n \times n$ image.
    We define a WEMD seminorm in terms of its discrete wavelet transform,
    \begin{align}
        \|\I\|_{\text{WEMD}}
        &:=
        \mathrm{WEMD} \left(\I, \bar{\I} \right)
        =
        \sum_{\ell, k, s_x, s_y}
        2^{-2\ell}
        \big|
            (\mathcal{W}\I)[\ell, k, s_x, s_y]
        \big| , \label{eq:WEMD_seminorm}
    \end{align}
    where $\bar{\I}$ is the \emph{mean image}, an image with all pixel values set to the mean pixel value of $\I$.
\end{definition}

\begin{figure}
    \hspace{-0.3cm}
    \begin{subfigure}{0.48\textwidth}
        \includegraphics[width=1.06\linewidth]{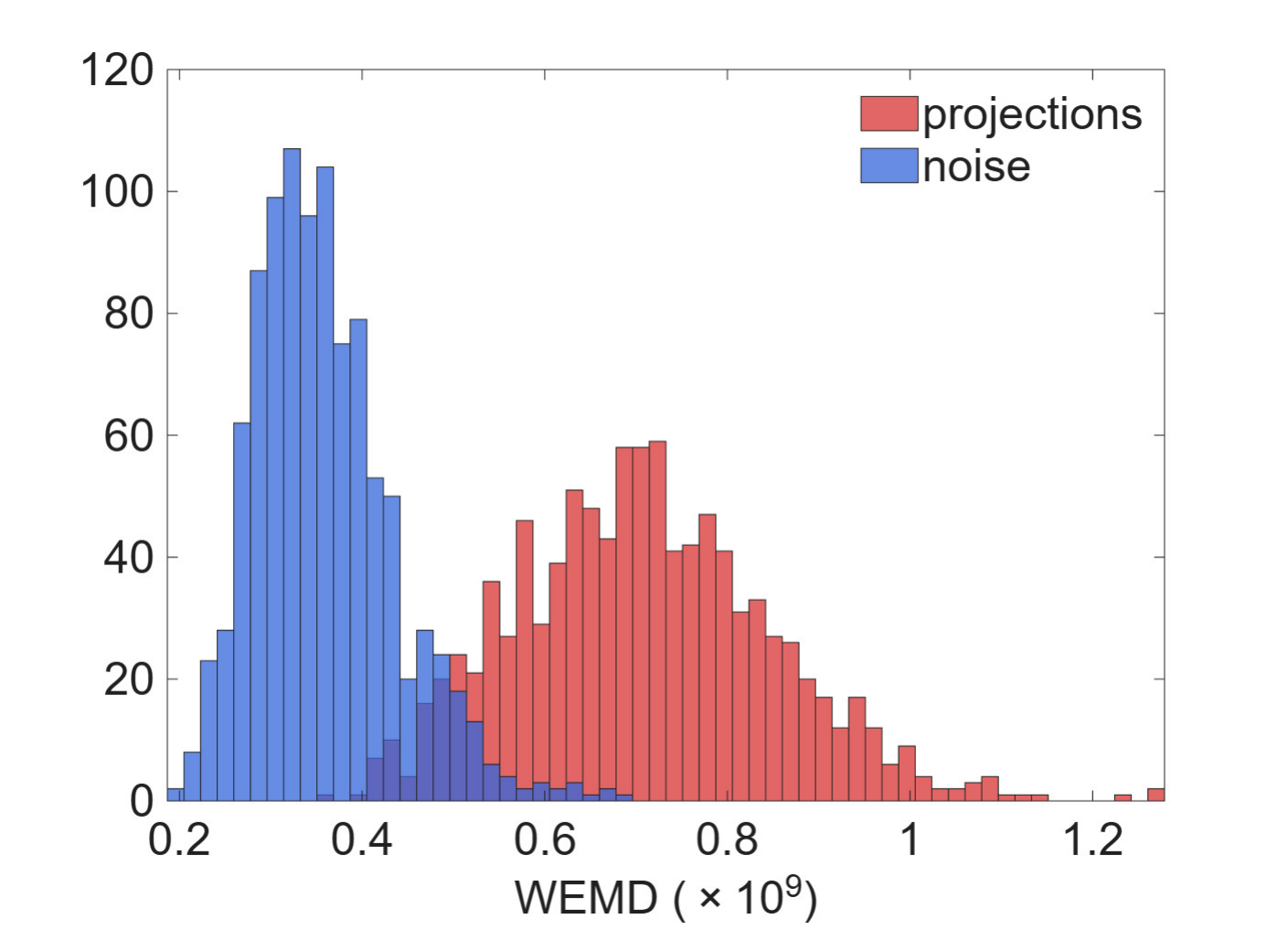}
        \caption{}
        \label{subfig:emd}
    \end{subfigure}
    \hspace{0.4cm}
        \begin{subfigure}{0.48\textwidth}
        \includegraphics[width=1.06\linewidth]{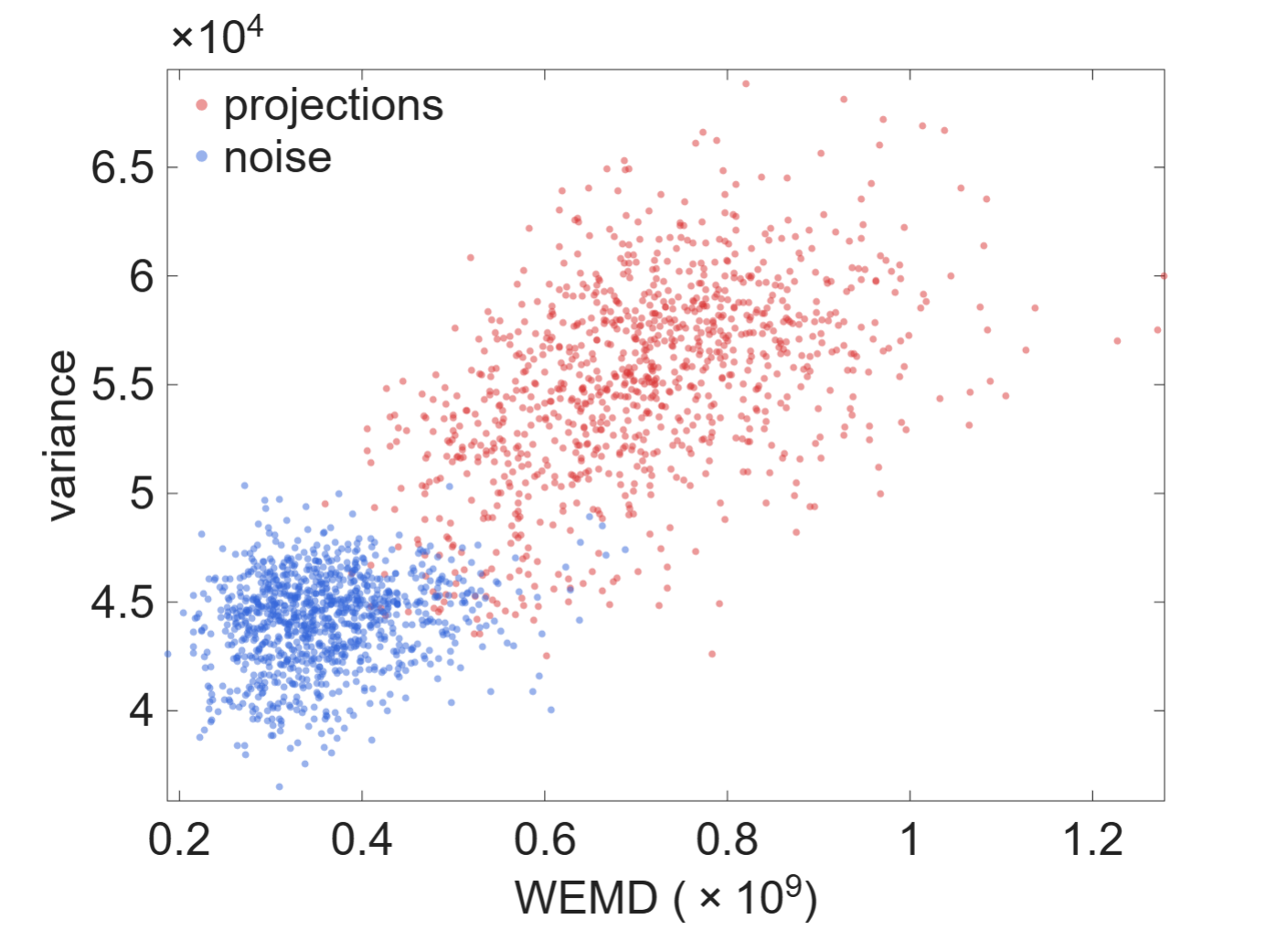}
        \caption{}
        \label{subfig:var}
    \end{subfigure}
    \caption{Comparison of projections and noise. We extracted $1000$ projection windows and $1000$ noise windows from the EMPIAR-10028 dataset \cite{wong2014cryo}. Projection windows were extracted according to coordinates included in the dataset and noise windows were extracted from regions that, according to the particle picking provided with the dataset, do not contain projections. (a) Histogram of the WEMD seminorm.
     (b) Empirical variance vs. WEMD seminorm.}
    \label{fig:fp_comparison}
\end{figure}

Since the DC coefficient is not included in the sum \eqref{eq:WEMD_seminorm}, this seminorm is valid even for images with negative values. It is nothing but the WEMD between an image and the same image where all non-DC coefficients have been zeroed. This is a surrogate for the $W_1$ transport cost of ``smoothing out'' an image.
Equivalently, it can be viewed as the approximate cost of transporting the positive pixels of $\I-\bar{\I}$ to the negative pixels.
In pure noise images, pixels above the mean usually lie near pixels below it, so the excess mass only has to move short distances to smooth out the noise.
By the definition of the earth mover's distance, moving mass between nearby pixels has a low associated cost, 
and so transporting the mass of a pure noise image to its mean image typically has a low overall cost. 
This phenomenon was observed empirically \cite{KileelMoscovichZeleskoSinger2021, SingerYang2024}
and was recently analyzed by Lager et al.~\cite{LagerMordantMoscovich2026}. 
In contrast, when the image contains a particle projection, it typically has large contiguous areas with either low or high values.
In this case, to transport $\I$ to $\bar{\I}$, masses must be transferred over large distances, resulting in high overall costs.
Figure~\ref{subfig:emd} shows histograms of the values of $\|\I\|_{\text{WEMD}}$ for pure noise windows vs. noisy particle projections, demonstrating good separation.

\begin{remark}
A seminorm similar to the WEMD seminorm of Eq.~\eqref{eq:WEMD_seminorm} but based on the exact earth mover's distance has been studied by Puthawala et al.~\cite{PuthawalaHauckOsher2019} under the name \emph{structure}.
\end{remark}

\section{Method description} \label{sec:method}

In this section we present our WEMD-based particle picker. It is a binary classifier, similar to other pickers \cite{wang2016deeppicker, heimowitz2018, bepler2019positive}. However, while Wang et al.~\cite{wang2016deeppicker} use micrographs from other molecules for training and Bepler et al.~\cite{bepler2019positive} use human-annotated positive examples, our method is self-supervised, using the WEMD seminorm to automatically select positive and negative examples of the molecule of interest from the micrograph.

\subsection{Self-supervised selection of training set}
\label{subsubsec:training}

It is well established that areas of the micrograph that contain a tomographic projection have a lower mean intensity and higher variance than areas of pure noise\footnote{This is prior to CTF correction.}~\cite{vanheel1982, nicholson2009}. Therefore, our training set is a set of vectors $\mathbf{x}_1, \dots, \mathbf{x}_M \in \mathbb{R}^2$ and  their associated binary classifications $y_1,\dots,y_M \in \{0,1\}$, where each vector $\mathbf{x}_i = \begin{bmatrix} \mu_i & \sigma_i \end{bmatrix}^T$ consists of the mean intensity and empirical standard deviation of an $n \times n$ window in the micrograph. Note that $n$ is chosen to make the windows slightly smaller than the size of a projection. 

To automatically select the training set, we begin by extracting many windows $\{g_i \}_{i=1}^{N}$, where $N \gg M$, from the micrograph. Each of these windows has a size of $n \times n$. To ensure that the majority of the support of each projection will be contained in one of the extracted windows, we choose windows on a grid with step size of $\nicefrac{n}{2}$~\cite{heimowitz2018}. That is, starting with the top left corner of the micrograph, we extract all windows with step size $\nicefrac{n}{2}$ until we reach the bottom right corner.

Next, we aim to select positive and negative examples of projections from our extracted windows. The windows with the highest (lowest) WEMD seminorm score are very good candidates for projection (pure noise) images. We therefore select the $\tau_1$ windows with highest WEMD seminorms, compute their mean and standard deviation and create $\mathbf{x}_1,\dots,\mathbf{x}_{\tau_1}$. Similarly, we select the $\tau_2$ windows with lowest WEMD seminorms, compute their mean and standard deviation and create $\mathbf{x}_{\tau_1+1},\dots,\mathbf{x}_{M=\tau_1+\tau_2}$. Together with their known labels, the training set is complete.

\begin{remark}
    In micrographs with large variations in ice thickness we recommend using low values of $\tau_1$ and $\tau_2$. The reason for this is that when there is a large variation in ice thickness, the variance of the noise changes accordingly. That is, thicker ice results in fewer electrons being elastically scattered by the molecule and fewer electrons contributing to the shot noise~\cite{ShiSinger2022}. This means that the seminorm of projection images in thin ice may have similar seminorm to noise windows in thick ice regions. To avoid training our model on incorrect examples, it is best to keep the training set small.
\end{remark}

\subsection{SVM classification and segmentation}
\label{subsubsec:svm}

\begin{figure*}
    \centering
    \begin{subfigure}{0.3\textwidth}
        \includegraphics[width=\linewidth]{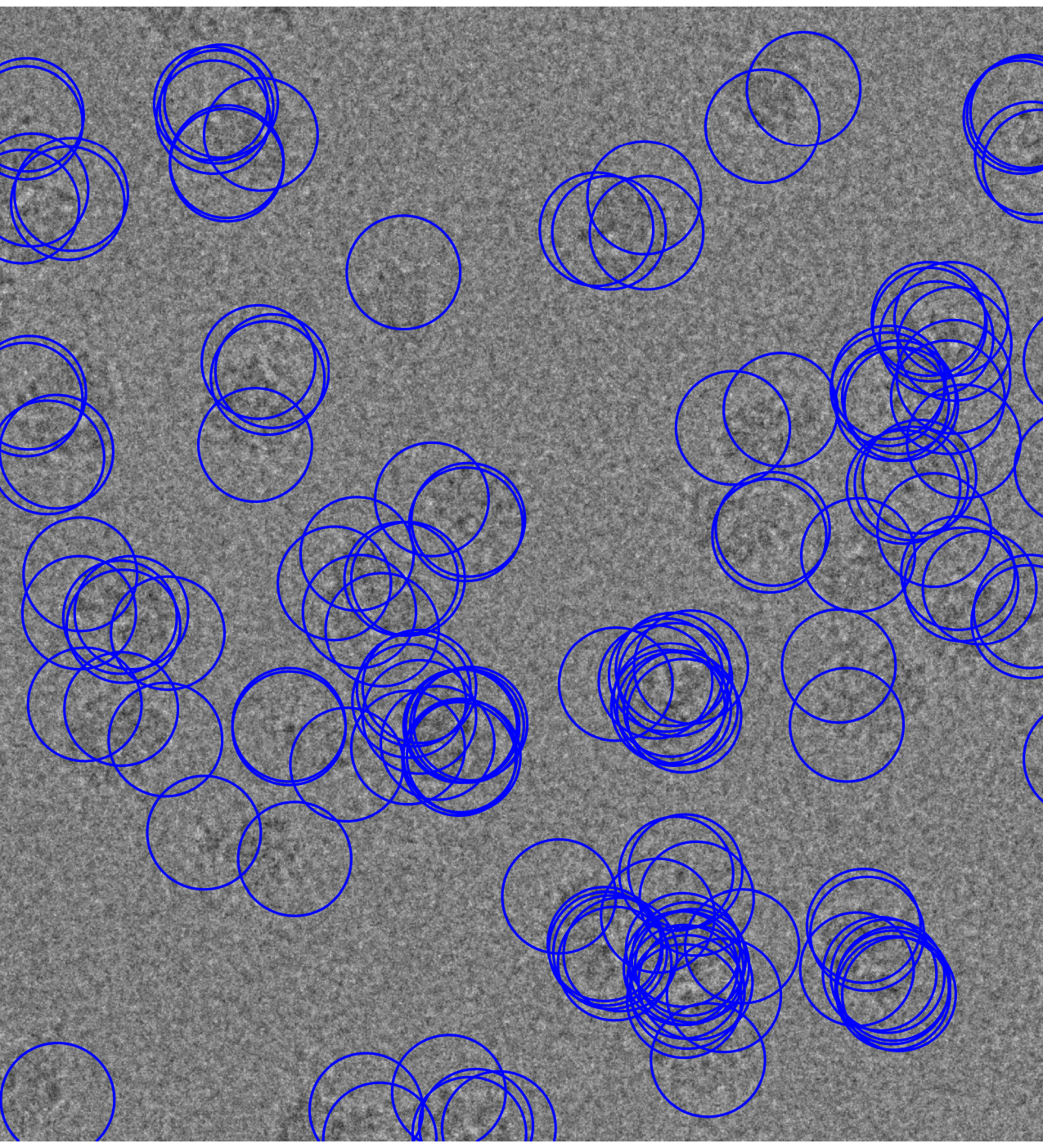}
        \caption{}
        \label{fig:before_centring}
    \end{subfigure}
    \begin{subfigure}{0.3\textwidth}  
        \includegraphics[width=\linewidth]{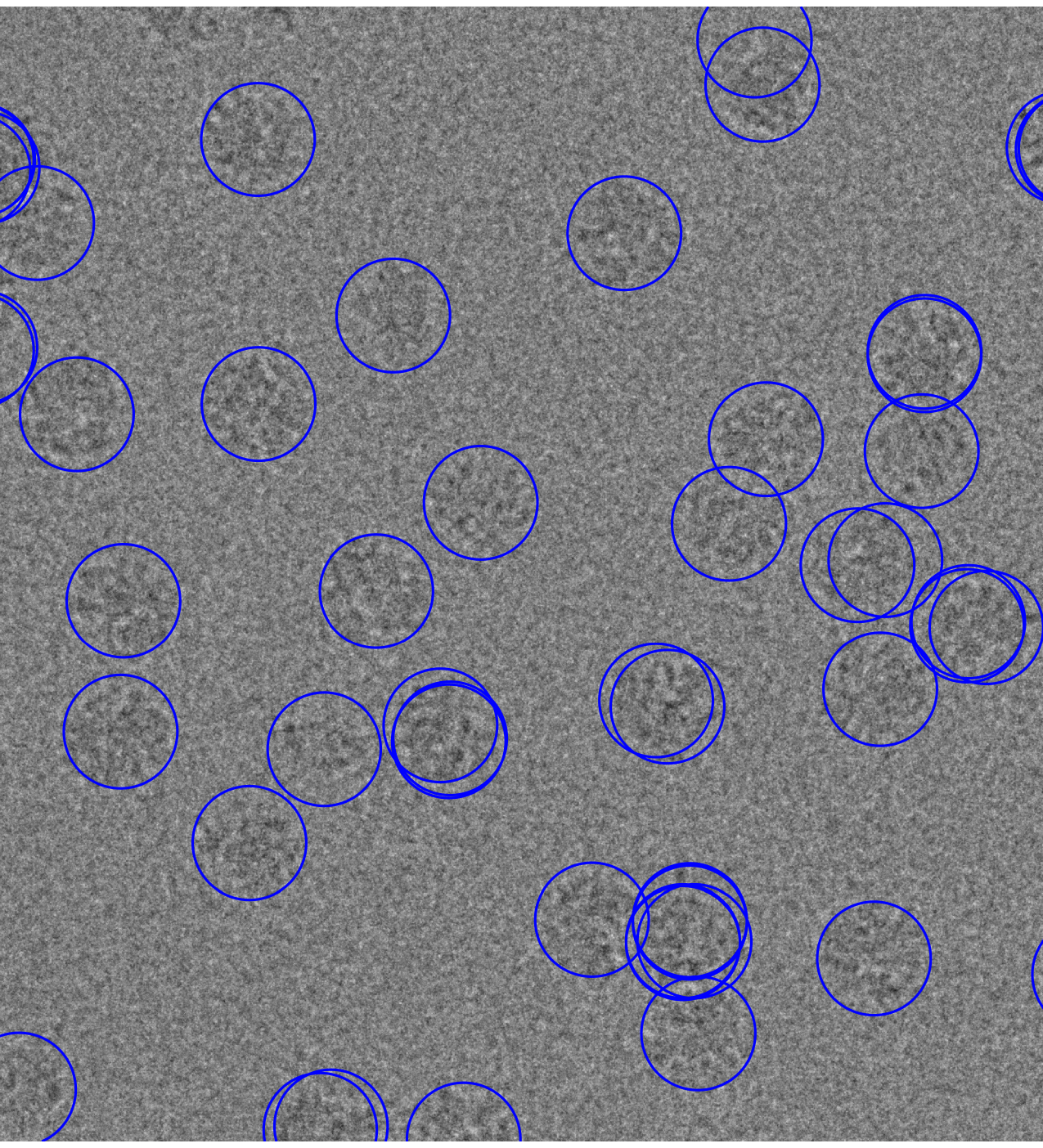}
        \caption{}
        \label{fig:after_centering}
    \end{subfigure}
    \begin{subfigure}{0.3\textwidth}  
        \includegraphics[width=\linewidth]{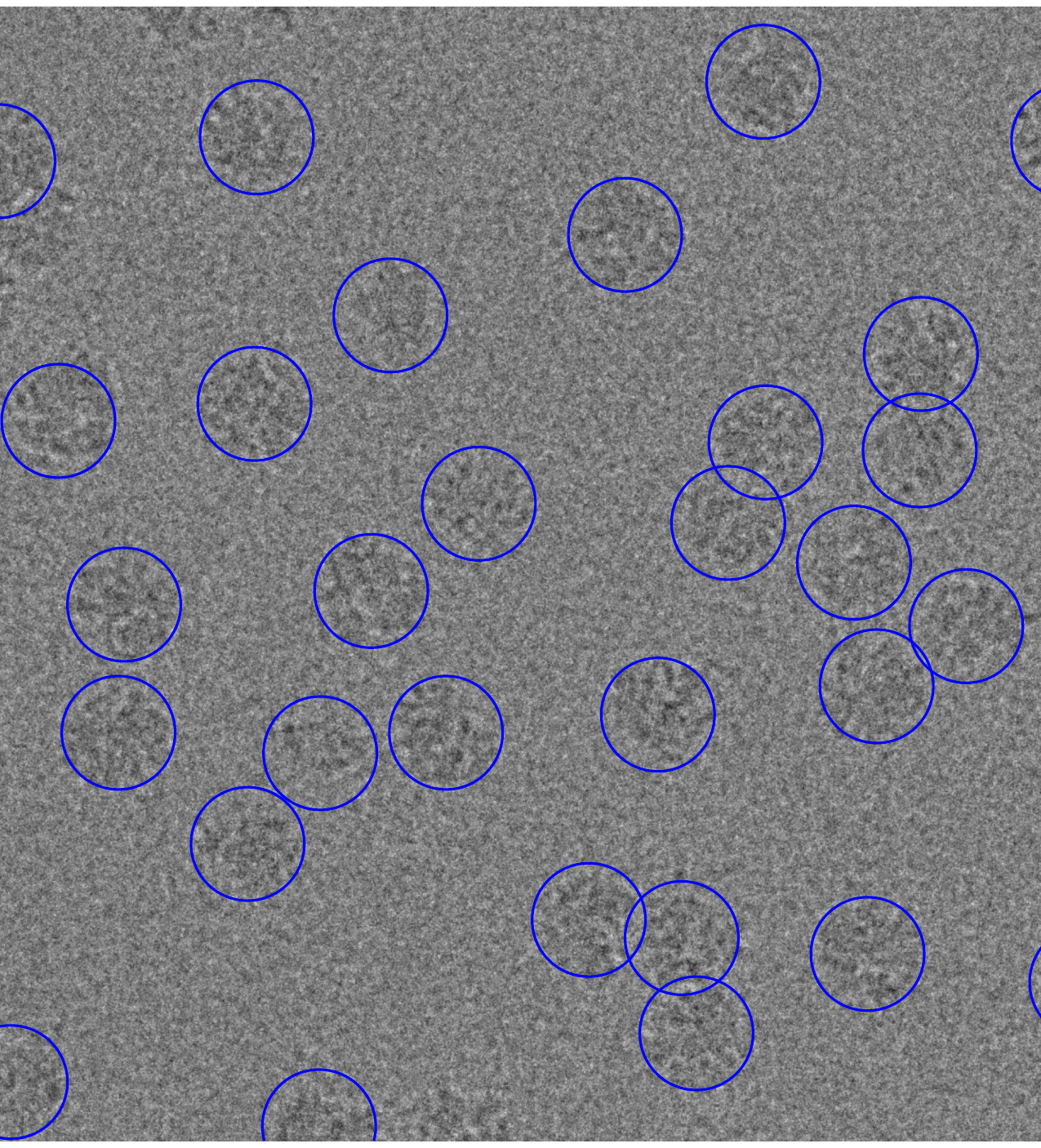}
        \caption{}
        \label{fig:overlap_filtering}
    \end{subfigure}
    \caption{Particle picking results. (a) Initial particle picking, including multiple picks per projection. (b) Interim particle picking, after centering each initial pick. (c) Final particle picking.}
    \label{fig:centering_comparison}
\end{figure*}

We use the automatically selected training set to train an SVM classifier on the mean and standard deviation cues, as suggested in~\cite{heimowitz2018}. 
Once the classifier has been trained, we use it to classify each possible $n \times n$ window in the micrograph and assign the classification to the central pixel in each window. In this manner, we produce a segmentation of the micrograph into projection pixels and noise pixels. In this segmentation, denoted as $I_{S}$, projection pixels are arranged in clusters. This is due to the fact that neighboring pixels in the segmentation are classified according to the mean and standard deviation of windows with a significant overlap. Our initial picked particle locations are given by the central pixel of each cluster.
The size of the clusters is determined by the size of the training set. When the training set includes many positive (projection) examples, some vectors $\mathbf{x}_1,\dots,\mathbf{x}_{\tau}$ must be the mean and standard deviation of windows that contain a small portion of the projection. As a consequence, when computing $I_{S}$, pixels on and around the support of a projection will be classified as projection, causing projections that are spaced close together to be considered part of the same cluster. This leads to loss of particles and bad localization of projections.
On the other hand, when the training set includes few positive (projection) examples, then a single projection may be represented as several clusters in $I_S$ (an illustration of this issue is presented in Figure~\ref{fig:before_centring}). This issue is easily fixed. First, we remove clusters that are much larger than the size of a projection as these are likely artifacts such as the carbon film. For the remaining clusters, we identify the central pixel in each cluster and extract a window around this pixel as a possible projection window (see Figure~\ref{fig:before_centring}). We then estimate the center of mass of each projection using the method of~\cite{heimowitz2020centering}. This procedure allows us to eliminate many clusters as in most cases, all clusters sharing a projection will converge to the same estimate of the center of mass. We use these centers of mass as our interim picked particle locations (for an illustration see Figure~\ref{fig:after_centering}). Lastly, we eliminate picks that remain too close (see Figure~\ref{fig:overlap_filtering}).

\begin{remark}
    The WEMD seminorm depends, among other factors, on the noise variance. Namely, a {significant} growth in noise variance will typically lead to transport of larger mass and, therefore, larger WEMD.  We present a scatter plot of variance vs. WEMD in Figure~\ref{subfig:var}. 

    Importantly, the WEMD seminorm is not used as a feature for the SVM classifier and variance is not used for training set selection as this may lead to  overfitting. Specifically, if the variance of the training set is extremely high for positive examples and extremely low for negative examples, the classifier will likely not generalize well. The same is true for the seminorm.
\end{remark}

\section{Theory} \label{sec:theory}

In this section we present a mathematical analysis of the WEMD seminorm (Definition~\ref{def:wemd_seminorm}) under a simplified i.i.d. white noise model that explains why one would expect it to separate pure noise windows from particle windows. 
This analysis assumes that the DWT is an orthogonal transformation whose output dimension equals the input dimension.\footnote{Not every DWT is orthogonal. When using the ``periodization'' mode of PyWavelets  with  an orthogonal wavelet family on image sizes that are powers of two the DWT is indeed orthogonal and dimension-preserving. However, in general there may be boundary padding or other effects due to rescaling that result in transformations that are not strictly orthogonal, or whose output dimension is slightly larger than the input dimension.}

First, we note the following fact about orthonormal transformations of i.i.d. Gaussians:
\begin{lemma} \label{lemma:gaussian_dwt_invariance}
    Let $\M \in \R^{p \times p}$ be an orthonormal matrix and let $\x \sim \mathcal{N}(0, \sigma^2 \I_{p \times p})$ be a random vector with independent $\mathcal{N}(0, \sigma^2)$ entries.
    Then $\M \x \sim  \mathcal{N}(0, \sigma^2 \mathbf{I}_{p \times p})$.
\end{lemma}
\begin{proof}
    For any matrix $\M$, if $\x \sim \mathcal{N}(\boldsymbol{\mu}, \boldsymbol{\Sigma})$ then it follows that $\M \x \sim \mathcal{N}(\M \boldsymbol{\mu}, \M \boldsymbol{\Sigma} \M^T)$.
    In our case, $\boldsymbol{\mu}=0$ and $\Sigma = \sigma^2 I$.
    Therefore, $\M \x \sim \mathcal{N}(0, \sigma^2 \M \M^T)$ and since $\M$ is orthonormal, $\M \M^T = I$.
\end{proof}
Thus for an orthonormal wavelet transform $\mathcal{W}$, if the pixels of an image are drawn independently from $\mathcal{N}(0, \sigma^2)$ then the wavelet coefficients follow the same distribution.
In that case the sum in Eq.~\eqref{eq:WEMD_seminorm} is nothing but a weighted sum of absolute values of i.i.d. Gaussians.
Each absolute value is a \emph{folded normal distribution} with mean and variance given by
\begin{align}
    \mathbb{E} |\mathcal{N}(0, \sigma^2)|
    =
    \sqrt{\frac{2}{\pi}} \sigma,
    \quad
    \text{Var}{|\mathcal{N}(0, \sigma^2)|}
    =
    \left( 1 - \frac{2}{\pi} \right) \sigma^2.
\end{align}
For simplicity, we shall assume that the image size is a power of two and that the level $L$ gives a full dyadic decomposition such that $n = 2^L$.
Under those assumptions, a 2-D DWT with no boundary padding has exactly $2^\ell \times 2^\ell$ coefficients for every level $\ell \in \{0, \ldots, L-1\}$ and every $k \in \{1,2,3\}$.
By linearity of expectation we obtain the expected WEMD seminorm of a pure noise patch,
\begin{align}
    \mathbb{E} \left[ \| \I \|_{\mathrm{WEMD}} \right]
    &=
    \sum_{\ell=0}^{L-1}
    2^{-2\ell}
    \sum_{k}
    \sum_{s_x, s_y}
    \mathbb{E}
    \big|
        (\mathcal{W}\I)[\ell, k, s_x, s_y]
    \big| \\
    &=
    \sum_{\ell=0}^{L-1}
    2^{-2\ell}
    \cdot
    3 \cdot 2^{2\ell} \cdot \mathbb{E} |\mathcal{N}(0, \sigma^2)|
    = \frac{3\sqrt{2}}{\sqrt{\pi}} L \sigma  \label{eq:E_I_WEMD_result}
\end{align}
where $L = \log_2(n)$.
Furthermore, by Lemma~\ref{lemma:gaussian_dwt_invariance} the wavelet coefficients under the i.i.d. Gaussian setting are independent, hence the variance is simply the sum of variances,
\begin{align}
    \mathrm{Var} \left( \| \I \|_{\mathrm{WEMD}} \right)
    &=
    \sum_{\ell=0}^{L-1} \left( 2^{-2\ell} \right)^{2}
    \sum_{k}
    \sum_{s_x, s_y}
    \mathrm{Var}\,\big| (\mathcal{W}\I)[\ell, k, s_x, s_y] \big| 
    \\
    &=
    \sum_{\ell=0}^{L-1} 2^{-4\ell}
    \cdot 3 \cdot 2^{2\ell}
    \cdot \mathrm{Var}\,\big| \mathcal{N}(0, \sigma^2) \big|
    =
    4\left( 1 - \tfrac{2}{\pi} \right)\left( 1 - 4^{-L} \right)\sigma^{2}.
\end{align}
For large values of $L$ this variance is approximately $1.45 \sigma^2$.
To summarize, under the pure noise model with pixel values drawn from $\mathcal{N}(0, \sigma^2)$ independently,
\begin{align} \label{eq:IWEMD_noisepatch_summary}
    \| \I \|_{\mathrm{WEMD}} \approx 2.4 \log_2(n) \sigma &\pm 1.2 \sigma.\\
    \text{(expectation} &\pm \text{stdev)} \nonumber
\end{align}
Let us now consider the alternative hypothesis that the image contains a signal, $\I = s + \eta$, where $s$ is the clean projection $h * P(T v)$ of Eq.~\eqref{equ:image_formation} and $\eta$ is i.i.d.\ $\mathcal{N}(0, \sigma^2)$ noise.
We define the signal-to-noise ratio of the window\footnote{Note that this SNR is different from the SNR of the whole micrograph.} as
\begin{align} \label{eq:SNR}
    \mathrm{SNR} := \frac{\|s\|^2}{n^2 \sigma^2}.
\end{align}
An important property of wavelet decompositions is that they have fast coefficient decay for various types of natural signals. In other words, the energy of typical images is concentrated in a handful of wavelet coefficients \cite{Mallat1989, Simoncelli1999}. 
For example, consider an extreme case where the energy of the signal is spread evenly throughout the wavelet coefficients at level $\ell = 1$.
In that case all the energy of $s$ is divided equally among the $3 \cdot 2^{2} = 12$ wavelet coefficients $c_1, \ldots, c_{12}$ of this level.
By orthogonality, for those coefficients
\begin{align}
    |c_i|^2
    =
    \tfrac{1}{12} \|\mathcal{W}s\|^2
    =
    \tfrac{1}{12} \|s\|^2
    =
    \tfrac{1}{12} \mathrm{SNR} \cdot  n^2 \sigma^2.
\end{align}
Thus,
\begin{align}
    \| s \|_{\mathrm{WEMD}}
    =
    2^{-2} \sum_{i=1}^{12} |c_i|
    =
    \frac{\sqrt{3}}{2} \cdot \sqrt{\mathrm{SNR}} \cdot n \sigma.
\end{align}
For a fixed particle image, adding the noise raises the expectation of the WEMD seminorm by roughly $2.4 \log_2(n)\sigma  - 2.4 \sigma$, while the standard deviation remains $1.2\sigma$.
The reason we subtract $2.4 \sigma$ from the expectation is that the coefficients at all levels other than $\ell = 1$ are pure noise and contribute $2.4 \sigma$ per level, as in Eq.~\eqref{eq:E_I_WEMD_result}, whereas each of the $12$ coefficients at level $\ell = 1$ becomes $c_i + z$ with $z \sim \mathcal{N}(0, \sigma^2)$.
For large coefficients that satisfy $|c_i| \gg \sigma$ we have $\mathbb{E}|c_i + z| \approx |c_i|$ so we need to discount the contribution of the noise at that level.
Hence, the mean gap between particle windows and pure noise windows, in standard deviations, is roughly
\begin{align} \label{eq:effect_size_in_sigmas}
    \frac
    {\left(
        \tfrac{\sqrt{3}}{2} \cdot \sqrt{\mathrm{SNR}} \cdot n \sigma
        +
        2.4(\log_2(n) - 1)\sigma
    \right)
    -
        2.4 \log_2(n) \sigma
    }
    {1.2\sigma}
    \approx
    0.72 \sqrt{\mathrm{SNR}} \cdot n - 2.
\end{align}
A typical SNR for particle windows in cryo-EM is $0.001$. Plugging this into Eq.~\eqref{eq:effect_size_in_sigmas} with $n=200$ gives $2.55$ standard deviations which suggests good separation.

\section{Experimental Results}
\label{sec:exp}

\subsection{Real-world datasets}

To demonstrate our suggested self-supervised particle picker, we used three  datasets from the Electron Microscopy Public Image Archive (EMPIAR)~\cite{iudin2023empiar}: a dataset of the 80S ribosome~\cite{wong2014cryo} known as EMPIAR-10028, a dataset of the T20S proteasome~\cite{danev2016cryo} known as EMPIAR-10057 and a dataset of {$\beta$}-galactosidase~\cite{scheres2015} known as EMPIAR-10017. We selected these datasets because they include the coordinates of picked particles that can be compared to our picked particles to verify the success of our method. Specifically, we computed two 3-D reconstructions for each of the datasets. The first reconstruction was done from projection coordinates included in the dataset entry and the second reconstruction was done from coordinates selected by our particle picker. We show that the reconstruction from our coordinates is comparable to the reconstruction from the original coordinates.

\subsubsection{80S ribosome}
\label{subsubsec:10028}

The EMPIAR-10028 dataset consists of $1081$ motion-corrected micrographs, as well as the CTF parameters and picked particle coordinates for each micrograph. For each of these micrographs, we use our suggested method to pick particles. 
We first extract all windows of size $220 \times 220$ with a stride of $110$. We then compute the WEMD seminorm of each extracted window. Since the windows with highest WEMD, denoted $g_1,\dots,g_{26}$, are most likely to contain a projection, we use these windows to create positive examples for our classifier. We use the $26$ windows with lowest WEMD to create negative examples. 

Once the training set has been selected and the SVM classifier has been trained, we perform particle picking as detailed in Section~\ref{subsubsec:svm}. Initially, our picker selected $182,143$ particles, while ensuring that between any two particle coordinates there is a distance of at least $120$ pixels ($160$~\AA).\footnote{The coordinates included in the dataset entry have a lower minimal distance than $120$ pixels.} The runtime of our method was on average $3.6$ seconds per micrograph when running on an Intel Core Ultra 7 CPU with 32 GB RAM and NVIDIA RTX 2000 GPU, and using a fast approximated centering method rather than the method of~\cite{heimowitz2020centering}.\footnote{Submitted for publication, but as yet unpublished.}

We then used RELION~\cite{zivanov2018new} to compute two reconstructions. The first reconstruction was computed using the coordinates provided with the dataset entry and the second reconstruction was done from our coordinates. For each reconstruction, we perform 2-D classification using $100$ classes, initial model building, 3-D classification into $4$ classes, refinement and post-processing using RELION's post-process task. The results are summarized in Table~\ref{tab:extraction_rate_10028}. Our method has fewer outliers and slightly improved resolution.

\begin{table}[ht!]
    \centering
    \renewcommand{\arraystretch}{1.2} 
    \begin{tabular}{ccc}
        & Original & Ours \\ \hline \hline
        \# coordinates & 158,212  & 182,143 \\ 
        \hline
        Runtime per micrograph & Unknown & 3.6 seconds\\ 
        \hline
        2-D classification & 96,410 (60\%)   &  112,322 (62\%) \\
        \hline
        Resolution & 3.77\AA   &  3.71\AA \\
        \hline      
    \end{tabular}
    \caption{Comparison of 3-D reconstruction for the EMPIAR-10028 dataset. Since no images were discarded during the 3-D classification, this step is not included.}
    \label{tab:extraction_rate_10028}
\end{table}

\subsubsection{T20S proteasome}

The EMPIAR-10057 dataset consists of 158 micrographs of T20S proteasome acquired using a phase plate, as well as projection coordinates for each micrograph. As before, we compute two reconstructions using RELION. One reconstruction is done from the coordinates included in the dataset and the second reconstruction is from our picked particles.

When running our suggested particle picking method, we first extract all windows of size $110 \times 110$ with a stride of $55$. We then compute the WEMD seminorm. The training set is determined as detailed in Section~\ref{subsubsec:10028}, with $30$ positive and $30$ negative examples. Minimal distance between picked particles was $100$ pixels. The runtime of our method was on average 2.75 seconds per micrograph. For the reconstruction, we skip 2-D classification,  perform initial model building and 3-D classification into $6$ classes as described in~\cite{danev2016cryo}. We then perform refinement using RELION and report all results in Table~\ref{tab:extraction_rate_10057}. Once again, our picking results in fewer outliers without harming resolution. 

\begin{table}[ht!]
    \centering
    \renewcommand{\arraystretch}{1.2} 
    \begin{tabular}{ccc}
        & {Original} & {Ours} \\ \hline \hline
        {\# coordinates} & {35,469} & {27,691} \\ 
        \hline
        Runtime per micrograph & Unknown & 2.75 seconds\\ 
        \hline
        {3-D classification} & {22,654 (64\%)}   &  {23,410 (85\%)} \\
        \hline 
        {Resolution} & {4.1\AA}   &  {4.1\AA} \\
        \hline      
    \end{tabular}
    \caption{{Comparison of 3-D reconstruction for the EMPIAR-10057 dataset.}}
    \label{tab:extraction_rate_10057}
\end{table}

\subsubsection{\texorpdfstring{$\beta$}{Beta}-galactosidase}

The EMPIAR-10017 dataset consists of $84$ micrographs of $\beta$-galactosidase, as well as manually selected projection coordinates for each micrograph. We once again compare a reconstruction from our coordinates with the reconstruction from the manually selected coordinates.

When running our suggested particle picking method, we extract all windows of size $78 \times 78$ with a stride of $39$. The positive examples are created using the $62$ windows with highest WEMD seminorm. The negative examples are taken from the $62$ windows with lowest WEMD. Minimal distance between picked particles was $70$ pixels. The runtime of our method was on average $2.5$  seconds per micrograph. As we used a small training set, the particle picking included only $38,846$ projection. This number can be increased by adding more positive examples, however this may hurt localization leading to larger runtimes for 2-D classification.

We use the reconstruction procedure detailed in Section~\ref{subsubsec:10028}. The reconstruction resolution is identical. 

Additionally, we compare our results to the state-of-the-art particle picker Topaz~\cite{bepler2019positive}. When running Topaz, two primary approaches are available. The first approach is to simply use a pretrained model for quick particle picking. The second approach is to fine-tune the pretrained model by providing a particle picking of the dataset of interest. We compare our results to the pretrained model. 
Reconstruction results are summarized in Table~\ref{tab:extraction_rate_10017}. 

\begin{table}[ht!]
    \centering
    \renewcommand{\arraystretch}{1.2}    
    \begin{tabular}{cccc}
        & Original & Topaz & Ours \\ \hline \hline
        \# coordinates &  40,863 & 45,000 & 38,846 \\ 
        \hline     
        Run time per micrograph & N/A & 4.4 seconds & 2.5 seconds\\
        \hline
        2-D classification & 35,201 (86\%)  & 37,324 (83\%) &  28,738 (74\%) \\
        \hline
        3-D classification & 29,814 (72\%)  & 32,017 (71\%) &  24,045 (62\%) \\
        \hline 
        Resolution & 4.36\AA   & 4.36\AA & 4.36\AA \\
         \hline 
    \end{tabular}
    \caption{Comparison of 3-D reconstruction for the EMPIAR-10017 dataset.}
    \label{tab:extraction_rate_10017}
\end{table}

\subsection{Synthetic dataset} \label{sec:synthetic_data}

\begin{figure*}
    \centering
    \begin{subfigure}{0.25\textwidth}
        \includegraphics[width=\linewidth]{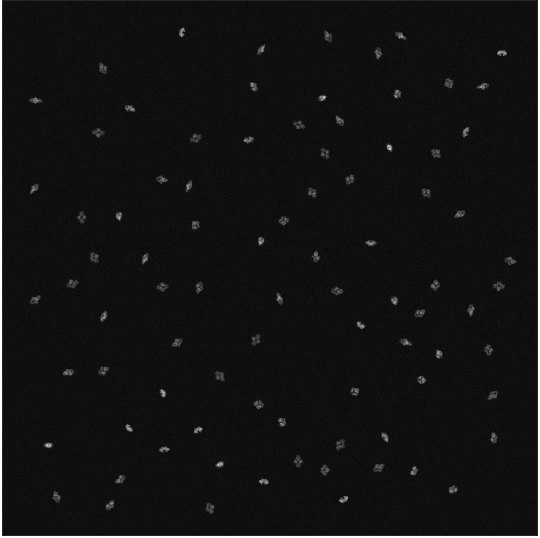}
        \caption{}
        \label{fig:c1}
    \end{subfigure}
    \begin{subfigure}{0.25\textwidth}
        \includegraphics[width=\linewidth]{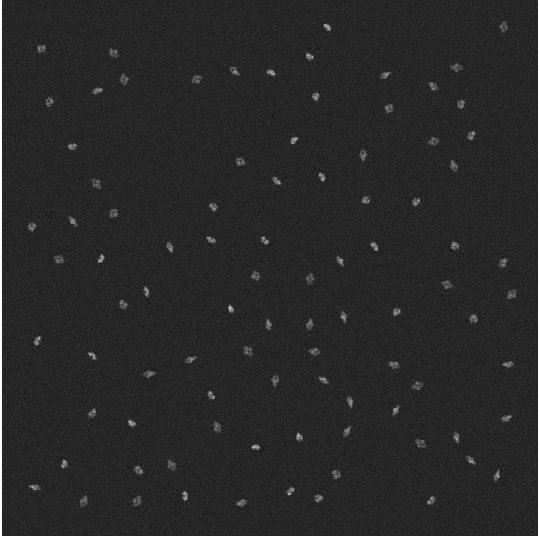}
        \caption{}
        \label{fig:c2}
    \end{subfigure}
    \begin{subfigure}{0.25\textwidth}
        \includegraphics[width=\linewidth]{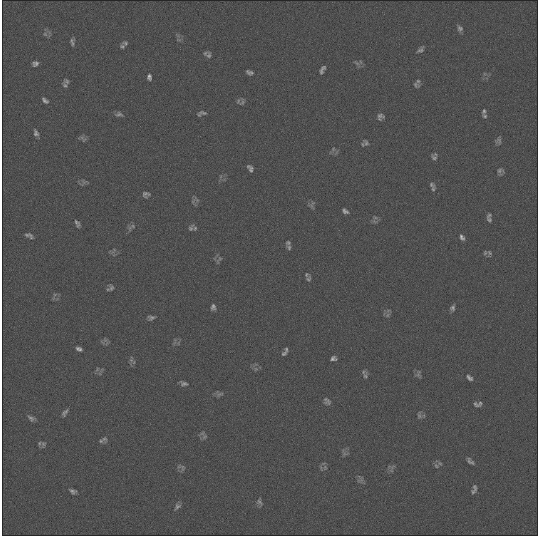}
        \caption{}
        \label{fig:c3}
    \end{subfigure}
    \begin{subfigure}{0.25\textwidth}
        \includegraphics[width=\linewidth]{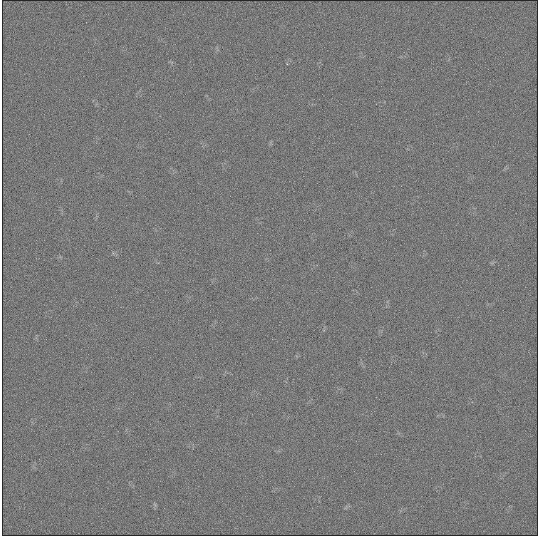}
        \caption{}
        \label{fig:c4}
    \end{subfigure}
        \begin{subfigure}{0.25\textwidth}
        \includegraphics[width=\linewidth]{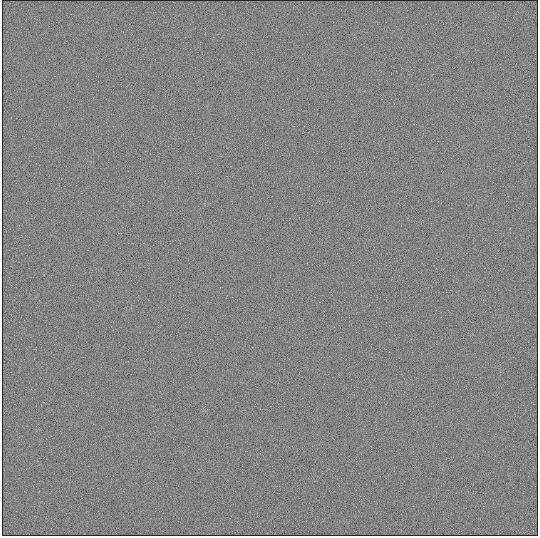}
        \caption{}
        \label{fig:c5}
    \end{subfigure}
    \caption{Synthetic micrograph examples. These micrographs were created from the volume EMD-2824 using the ASPIRE toolbox.}
    \label{fig:mex}
\end{figure*}

Since the real-world datasets do not have a ground truth, we created synthetic micrographs with varying levels of noise. We used the ASPIRE toolbox~\cite{garrett_wright_2025_17951406} to simulate $500$ micrographs. We first created projections of the 3-D volume EMD-2824 \cite{scheres2015} obtained from the Electron Microscopy Data Bank~\cite{wwPDB_EMDB_2024}. Each projection was taken in a random orientation. The projections were placed in random locations to create clean micrographs. Next, each micrograph was contaminated with additive i.i.d. Gaussian noise. We repeated this process several times to create micrographs where the windows' SNR (Eq.~\eqref{eq:SNR}) ranges from $0.00055$ to $0.0014$ on average over all projection windows of size $150 \times 150$ pixels. See Figure~\ref{fig:mex} for examples of micrographs with these levels of noise.

\begin{figure*}
    \centering
    \begin{subfigure}{0.45\textwidth}
        \includegraphics[width=\linewidth]{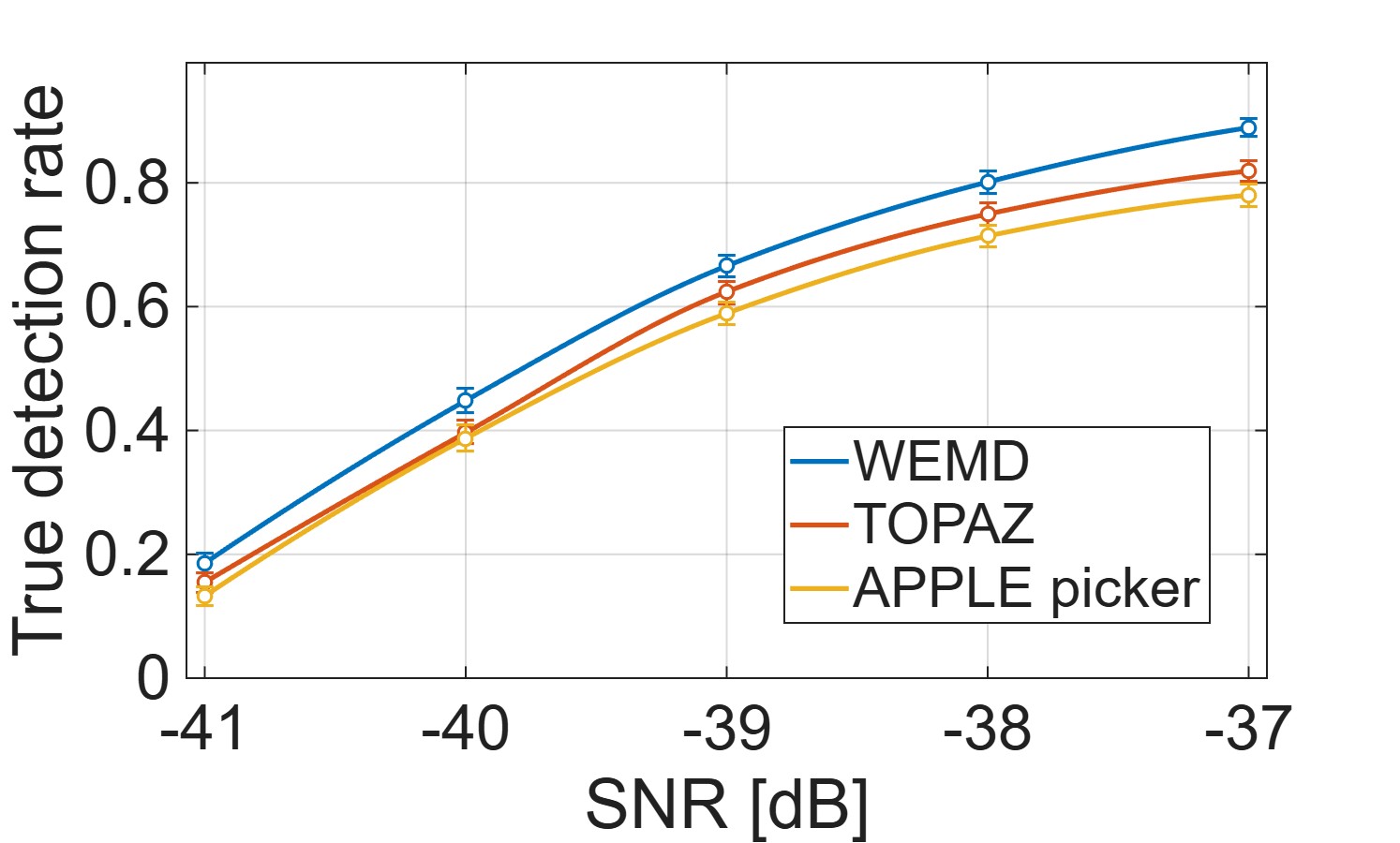}
        \caption{}
        \label{fig:td}
    \end{subfigure}
    \hspace{1cm}
    \begin{subfigure}{0.45\textwidth}
        \includegraphics[width=\linewidth]{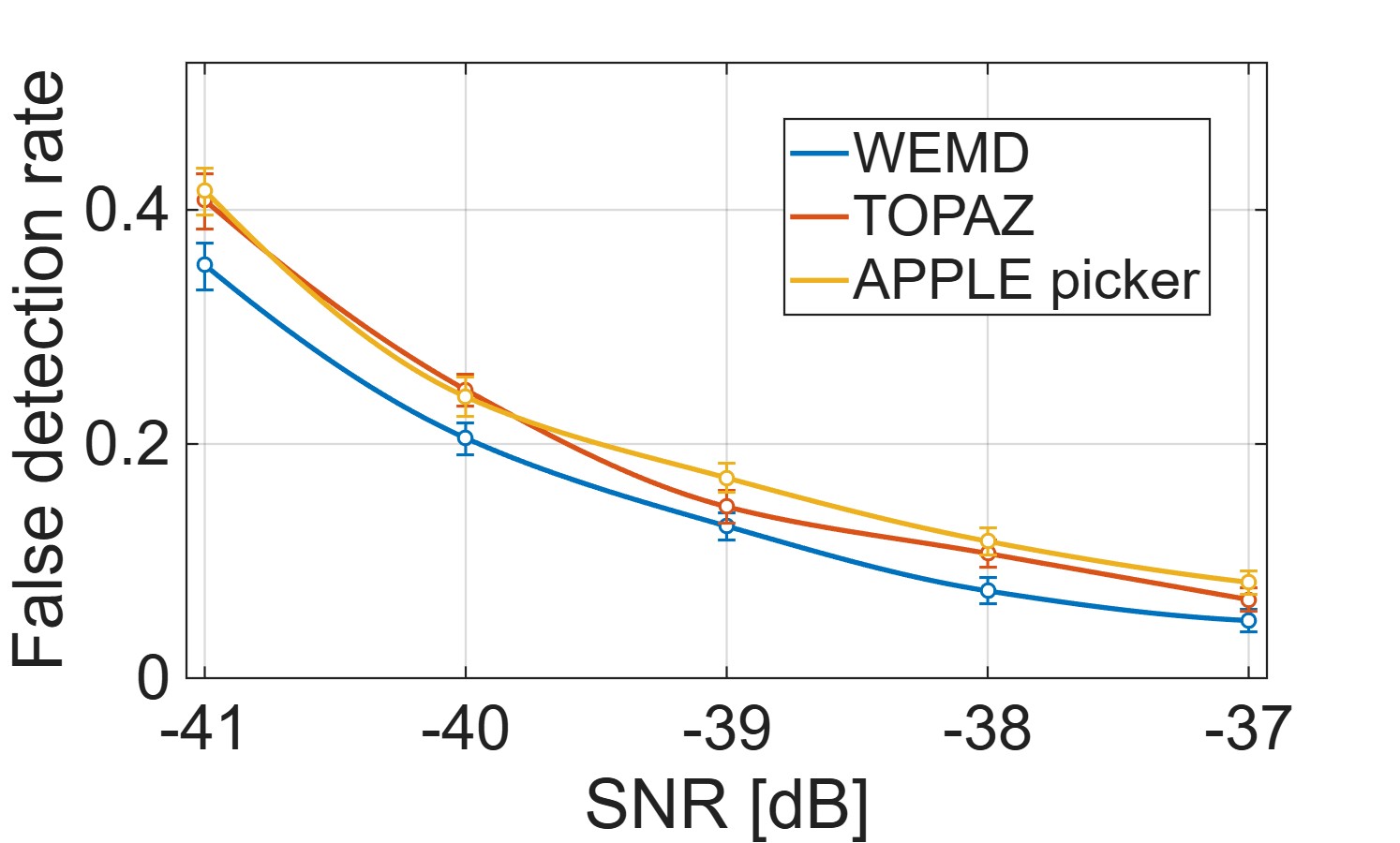}
        \caption{}
        \label{fig:fd}
    \end{subfigure}
    \caption{Results of synthetic experiment. (a) comparison of true detection rate between our suggested method, Topaz~\cite{bepler2019positive} and the APPLE-picker~\cite{heimowitz2018}. (b) comparison of false detection rate.}
    \label{fig:synthetic_graphs}
\end{figure*}

Using our suggested method, as well as the APPLE-picker~\cite{heimowitz2018} and Topaz~\cite{bepler2019positive}, we picked particles in each micrograph and computed the true detection and false detection rates. Our particle picker achieved a higher rate of true detection and a lower rate of false detections across all SNRs as compared to the APPLE-picker and Topaz \cite{heimowitz2018, bepler2019positive}. The results are summarized in Figure~\ref{fig:synthetic_graphs}.

\section{Conclusion}
\label{sec:con}
In this paper, we introduce a simple, fast, self-supervised  particle picker that identifies particle projections in a novel way using a wavelet earth mover's seminorm. 
The picker is evaluated on three real-world datasets as well as a synthetic dataset, where it demonstrated results comparable to state-of-the-art particle pickers. Specifically, when evaluated on the synthetic dataset, our particle picker consistently reached higher true detection and lower false detection rates than state-of-the-art particle pickers. 

\section*{Acknowledgements}

A.M. is supported in part by ISF Grant No. 1662/22 and NSF-BSF Grant No. 2022778.
The work of A.S. was supported in part by AFOSR FA9550-23-1-0249, the Simons Foundation Math+X Investigator Award, NSF DMS 2510039, and NIH/NIGMS R01GM136780-01.

The authors acknowledge the Ariel HPC Center at Ariel University for providing computing resources that have contributed to the research results reported in this paper.

Use of generative AI: Claude (Anthropic) assisted with code optimization, copy-editing and checking the mathematical derivations.

\bibliographystyle{unsrt}
\bibliography{WEMD_based_particle_picking}

@article{baker2010cryo,
  doi = {10.1038/nprot.2010.126},
  title={Cryo-{EM} of macromolecular assemblies at near-atomic resolution},
  author={Baker, Matthew L and Zhang, Junjie and Ludtke, Steven J and Chiu, Wah},
  journal={Nature Protocols},
  volume={5},
  number={10},
  pages={1697--1708},
  year={2010},
  publisher={Nature Publishing Group UK London}
}

@article{bai2015cryo,
  doi = {10.1016/j.tibs.2014.10.005},
  title={How cryo-{EM} is revolutionizing structural biology},
  author={Bai, Xiao-Chen and McMullan, Greg and Scheres, Sjors HW},
  journal={Trends in Biochemical Sciences},
  volume={40},
  number={1},
  pages={49--57},
  year={2015},
  publisher={Elsevier}
}

@article{BendoryBartesaghiSinger2019,
  doi = {10.1109/MSP.2019.2957822},
  title={Single-particle cryo-electron microscopy: {M}athematical theory, computational challenges, and opportunities},
  author={Bendory, Tamir and Bartesaghi, Alberto and Singer, Amit},
  journal={IEEE Signal Processing Magazine},
  volume={37},
  number={2},
  pages={58--76},
  year={2020},
}

@article{BhamreZhangSinger2016,
  doi = {10.1016/j.jsb.2016.04.013},
  title={Denoising and covariance estimation of single particle cryo-{EM} images},
  author = {Tejal Bhamre and Teng Zhang and Amit Singer},
  journal={Journal of Structural Biology},
  volume={195},
  number={1},
  pages={72--81},
  year={2016},
  publisher={Elsevier}
}

@article{bepler2019positive,
  doi = {10.1038/s41592-019-0575-8},
  title={Positive-unlabeled convolutional neural networks for particle picking in cryo-electron micrographs},
  author={Bepler, Tristan and Morin, Andrew and Rapp, Micah and Brasch, Julia and Shapiro, Lawrence and Noble, Alex J and Berger, Bonnie},
  journal={Nature Methods},
  volume={16},
  number={11},
  pages={1153--1160},
  year={2019},
}

@article{brilot2012,
title = {Beam-induced motion of vitrified specimen on holey carbon film},
journal = {Journal of Structural Biology},
volume = {177},
number = {3},
pages = {630-637},
year = {2012},
author = {Axel F. Brilot and James Z. Chen and Anchi Cheng and Junhua Pan and Stephen C. Harrison and Clinton S. Potter and Bridget Carragher and Richard Henderson and Nikolaus Grigorieff},
doi = {10.1016/j.jsb.2012.02.003},
}

@article{danev2016cryo,
  doi = {10.7554/eLife.13046},
  title={Cryo-{EM} single particle analysis with the Volta phase plate},
  author={Danev, Radostin and Baumeister, Wolfgang},
  journal={eLife},
  volume={5},
  pages={e13046},
  year={2016},
  publisher={eLife Sciences Publications, Ltd}
}

@article{eldar2020,
  doi = {10.1016/j.jsb.2020.107473},
title = {{KLT} picker: Particle picking using data-driven optimal templates},
journal = {Journal of Structural Biology},
volume = {210},
number = {2},
pages = {107473},
year = {2020},
issn = {1047-8477},
author = {Amitay Eldar and Boris Landa and Yoel Shkolnisky},
}

@book{Frank2006,
    author = {Frank, Joachim},
    title = {Three-Dimensional Electron Microscopy of Macromolecular Assemblies: Visualization of Biological Molecules in Their Native State},
    publisher = {Oxford University Press},
    year = {2006},
    month = {03},
    isbn = {9780195182187},
    doi = {10.1093/acprof:oso/9780195182187.001.0001}
}

@article{frank1983,
  doi = {10.1016/0304-3991(83)90256-5},
title = {Automatic selection of molecular images from electron micrographs},
journal = {Ultramicroscopy},
volume = {12},
number = {3},
pages = {169-175},
year = {1983},
author = {Joachim Frank and Terence Wagenknecht},
}

@article{grigorieff2011,
title = {Near-atomic resolution reconstructions of icosahedral viruses from electron cryo-microscopy},
journal = {Current Opinion in Structural Biology},
volume = {21},
number = {2},
pages = {265-273},
year = {2011},
author = {Nikolaus Grigorieff and Stephen C Harrison},
doi = {10.1016/j.sbi.2011.01.008},
}

@article{henderson2013avoiding,
  doi = {10.1073/pnas.1314449110},
  title={Avoiding the pitfalls of single particle cryo-electron microscopy: Einstein from noise},
  author={Henderson, Richard},
  journal={Proceedings of the National Academy of Sciences},
  volume={110},
  number={45},
  pages={18037--18041},
  year={2013},
  publisher={National Acad Sciences}
}

@article{henderson1992,
  doi = {10.1016/0304-3991(92)90003-3},
title = {Image contrast in high-resolution electron microscopy of biological macromolecules: TMV in ice},
journal = {Ultramicroscopy},
volume = {46},
number = {1--4},
pages = {1-18},
year = {1992},
issn = {0304-3991},
author = {Richard Henderson},
}

@article{heimowitz2020centering,
author = {Heimowitz, Ayelet and Sharon, Nir and Singer, Amit},
title = {Centering Noisy Images with Application to Cryo-{EM}},
journal = {SIAM Journal on Imaging Sciences},
volume = {14},
number = {2},
pages = {689-716},
year = {2021},
doi = {10.1137/20M1365946},
}

@article{heimowitz2018,
  doi = {10.1016/j.jsb.2018.08.012},
  title={{APPLE} picker: Automatic particle picking, a low-effort cryo-{EM} framework},
  author={Heimowitz, Ayelet and And{\'e}n, Joakim and Singer, Amit},
  journal={Journal of Structural Biology},
  volume={204},
  number={2},
  pages={215--227},
  year={2018},
  publisher={Elsevier}
}

@article{iudin2023empiar,
  doi = {10.1093/nar/gkac1062},
  title={{EMPIAR:} the {E}lectron {M}icroscopy {P}ublic {I}mage {A}rchive},
  author={Iudin, Andrii and Korir, Paul K and Somasundharam, Sriram and Weyand, Simone and Cattavitello, Cesare and Fonseca, Neli and Salih, Osman and Kleywegt, Gerard J and Patwardhan, Ardan},
  journal={Nucleic Acids Research},
  volume={51},
  number={D1},
  pages={D1503--D1511},
  year={2023},
  publisher={Oxford University Press}
}

@article{kapulin2025,
title = {Outlier removal in cryo-EM via radial profiles},
journal = {Journal of Structural Biology},
volume = {217},
number = {1},
pages = {108172},
year = {2025},
author = {Lev Kapnulin and Ayelet Heimowitz and Nir Sharon},
doi = {10.1016/j.jsb.2025.108172},
}

@INPROCEEDINGS{nguyen2018,
  doi = {10.1109/BIBM.2018.8621224},
  author={Nguyen, Nguyen P. and Ersoy, Ilker and White, Tommi and Bunyak, Filiz},
  booktitle={2018 IEEE International Conference on Bioinformatics and Biomedicine (BIBM)}, 
  title={Automated Particle Picking in Cryo-Electron Micrographs using Deep Regression}, 
  year={2018},
  volume={},
  number={},
  pages={2453-2460},
}

@article{nicholson2009,
  doi = {10.1006/jsbi.2001.4348},
title = {Review: Automatic Particle Detection in Electron Microscopy},
journal = {Journal of Structural Biology},
volume = {133},
number = {2--3},
pages = {90-101},
year = {2001},
author = {William V. Nicholson and Robert M. Glaeser},
}

@article{deoliveira2021,
  doi = {10.1177/2472555220960401},
title = {Cryo-EM: The Resolution Revolution and Drug Discovery},
journal = {SLAS Discovery},
volume = {26},
number = {1},
pages = {17-31},
year = {2021},
author = {Taiana Maia {de Oliveira} and Lotte {van Beek} and Fiona Shilliday and Judit E. Debreczeni and Chris Phillips},
}

@article{singer2020,
   author = "Singer, Amit and Sigworth, Fred J.",
   title = "Computational Methods for Single-Particle Electron Cryomicroscopy", 
   journal= "Annual Review of Biomedical Data Science",
   year = "2020",
   volume = "3",
   number = "1",
   pages = "163-190",
   publisher = "Annual Reviews",
   doi = "10.1146/annurev-biodatasci-021020-093826",  
  }

@article{shatsky2009,
title = {A method for the alignment of heterogeneous macromolecules from electron microscopy},
journal = {Journal of Structural Biology},
volume = {166},
number = {1},
pages = {67-78},
year = {2009},
author = {Maxim Shatsky and Richard J. Hall and Steven E. Brenner and Robert M. Glaeser},
doi = {10.1016/j.jsb.2008.12.008}
}

@article{scheres2015,
  doi = {10.1016/j.jsb.2014.11.010},
  title={Semi-automated selection of cryo-{EM} particles in {RELION}-1.3},
  author={Scheres, Sjors HW},
  journal={Journal of Structural Biology},
  volume={189},
  number={2},
  pages={114--122},
  year={2015},
  publisher={Elsevier}
}

@article{sigworth2004,
  doi = {10.1016/j.jsb.2003.10.025},
title = {Classical detection theory and the cryo-EM particle selection problem},
journal = {Journal of Structural Biology},
volume = {145},
number = {1--2},
pages = {111-122},
year = {2004},
author = {Fred J. Sigworth},
}

@article{vos2009,
  doi = {10.1016/j.jsb.2009.01.004},
title = {{D}o{G} {P}icker and {T}ilt{P}icker: {S}oftware tools to facilitate particle selection in single particle electron microscopy},
journal = {Journal of Structural Biology},
volume = {166},
number = {2},
pages = {205-213},
year = {2009},
issn = {1047-8477},
author = {N.R. Voss and C.K. Yoshioka and M. Radermacher and C.S. Potter and B. Carragher},
}

@article{vanheel1982,
title = {Detection of objects in quantum-noise-limited images},
journal = {Ultramicroscopy},
volume = {7},
number = {4},
pages = {331-341},
year = {1982},
author = {Marin {Van Heel}},
doi = {10.1016/0304-3991(82)90258-3},
}

@article{wang2016deeppicker,
  doi = {10.1016/j.jsb.2016.07.006},
  title={Deep{P}icker: {A} deep learning approach for fully automated particle picking in cryo-{EM}},
  author={Wang, Feng and Gong, Huichao and Liu, Gaochao and Li, Meijing and Yan, Chuangye and Xia, Tian and Li, Xueming and Zeng, Jianyang},
  journal={Journal of Structural Biology},
  volume={195},
  number={3},
  pages={325--336},
  year={2016},
}

@misc{garrett_wright_2025_17951406,
  author       = {Garrett Wright and
                  Joakim And{\'e}n and
                  Vineet Bansal and
                  Junchao Xia and
                  Chris Langfield and
                  Josh Carmichael and
                  Kris Sowattanangkul and
                  Robbie Brook and
                  Yunpeng Shi and
                  Ayelet Heimowitz and
                  Gabi Pragier and
                  Itay Sason and
                  Amit Moscovich and
                  Yoel Shkolnisky and
                  Amit Singer},
  title        = {{ASPIRE-Python}: v0.14.2},
  month        = dec,
  year         = 2025,
  publisher    = {Zenodo},
  version      = {v0.14.2},
  doi          = {10.5281/zenodo.17951406},
}

@article{wong2014cryo,
  doi = {10.7554/eLife.03080},
  title={Cryo-{EM} structure of the Plasmodium falciparum 80S ribosome bound to the anti-protozoan drug emetine},
  author={Wong, Wilson and Bai, Xiao-chen and Brown, Alan and Fernandez, Israel S and Hanssen, Eric and Condron, Melanie and Tan, Yan Hong and Baum, Jake and Scheres, Sjors HW},
  journal={eLife},
  volume={3},
  pages={e03080},
  year={2014},
  publisher={eLife Sciences Publications, Ltd}
}

@article{zivanov2018new,
  doi = {10.7554/eLife.42166},
  title={New tools for automated high-resolution cryo-{EM} structure determination in {RELION}-3},
  author={Zivanov, Jasenko and Nakane, Takanori and Forsberg, Bj{\"o}rn O and Kimanius, Dari and Hagen, Wim JH and Lindahl, Erik and Scheres, Sjors HW},
  journal={eLife},
  volume={7},
  pages={e42166},
  year={2018},
  publisher={eLife Sciences Publications, Ltd}
}

@article{KileelMoscovichZeleskoSinger2021,
    title={Manifold learning with arbitrary norms},
    volume={27},
    ISSN={1531-5851},
    doi={10.1007/s00041-021-09879-2},
    number={5},
    journal={Journal of Fourier Analysis and Applications},
    publisher={Springer Science and Business Media LLC},
    author={Kileel, Joe and Moscovich, Amit and Zelesko, Nathan and Singer, Amit},
    pages={82},
    year={2021},
    month=sep
}

@ARTICLE{zhang2017,
  doi = {10.1109/TCBB.2015.2415787},
  author={Zhang, Fa and Chen, Yu and Ren, Fei and Wang, Xuan and Liu, Zhiyong and Wan, Xiaohua},
  journal={IEEE/ACM Transactions on Computational Biology and Bioinformatics}, 
  title={A Two-Phase Improved Correlation Method for Automatic Particle Selection in Cryo-EM}, 
  year={2017},
  volume={14},
  number={2},
  pages={316-325},
}

@ARTICLE{wagner2019,
  doi = {10.1038/s42003-019-0437-z},
  author={Thorsten Wagner and Felipe Merino and Markus Stabrin and Toshio Moriya and  Claudia Antoni and Amir Apelbaum and Philine Hagel and Oleg Sitsel and Tobias Raisch and Daniel Prumbaum and Dennis Quentin and Daniel Roderer and Sebastian Tacke and Birte Siebolds and Evelyn Schubert and Tanvir R. Shaikh and Pascal Lill and Christos Gatsogiannis and Stefan Raunser},
  journal={Communications Biology}, 
  title={{SPHIRE}-cr{YOLO} is a fast and accurate fully automated particle picker for cryo-{EM}}, 
  year={2019},
  volume={2},
  number={1},
  pages={218}
}

@INPROCEEDINGS{shirdhonkar2008,
  author={Shirdhonkar, Sameer and Jacobs, David W.},
  booktitle={2008 IEEE Conference on Computer Vision and Pattern Recognition}, 
  title={Approximate earth mover’s distance in linear time}, 
  year={2008},
  volume={},
  number={},
  pages={1-8},
  doi={10.1109/CVPR.2008.4587662}}

@article{ShiSingerVerbeke2026,
    title={Fast Rigid Alignment of Heterogeneous Images in Sliced Wasserstein Distance},
    volume={19},
    ISSN={1936-4954},
    DOI={10.1137/25m1742916},
    number={3},
    journal={SIAM Journal on Imaging Sciences},
    publisher={Society for Industrial & Applied Mathematics (SIAM)},
    author={Shi, Yunpeng and Singer, Amit and Verbeke, Eric J.},
    year={2026},
    month=jul,
    pages={1469–1515}
  }

@article{PuthawalaHauckOsher2019,
    title={Diagnosing Forward Operator Error Using Optimal Transport},
    volume={80},
    ISSN={1573-7691},
    DOI={10.1007/s10915-019-00989-0},
    number={3},
    journal={Journal of Scientific Computing},
    publisher={Springer Science and Business Media LLC},
    author={Puthawala, Michael A. and Hauck, Cory D. and Osher, Stanley J.},
    year={2019},
    month=jul,
    pages={1549--1576}
}

@inproceedings{Simoncelli1999,
    title = {Modeling the joint statistics of images in the wavelet domain},
    booktitle = {{{SPIE}}'s {{International Symposium}} on {{Optical Science}}, {{Engineering}}, and {{Instrumentation}}},
    author = {Simoncelli, Eero P.},
    editor = {Unser, Michael A. and Aldroubi, Akram and Laine, Andrew F.},
    volume = {3813},
    year = 1999,
    month = oct,
    pages = {188--195},
    address = {Denver, CO},
    doi = {10.1117/12.366779},
    urldate = {2025-09-02},
    langid = {english}
}

@article{Mallat1989,
    title = {A theory for multiresolution signal decomposition: the wavelet representation},
    shorttitle = {A theory for multiresolution signal decomposition},
    author = {Mallat, S.G.},
    year = 1989,
    month = jul,
    journal = {IEEE Transactions on Pattern Analysis and Machine Intelligence},
    volume = {11},
    number = {7},
    pages = {674--693},
    doi = {10.1109/34.192463},
    urldate = {2025-11-10}
}

@article{CraigYu2025,
  title = {Wavelet s-{{Wasserstein}} distances for $0 < s \leq 1$},
  author = {Craig, Katy and Yu, Haoqing},
  year = 2025,
  month = aug,
  journal = {Sampling Theory, Signal Processing, and Data Analysis},
  volume = {23},
  number = {2},
  pages = {19},
  doi = {10.1007/s43670-025-00113-4},
  urldate = {2025-08-11},
  langid = {english}
}

@article{wwPDB_EMDB_2024,
  author       = {{wwPDB Consortium}},
  title        = {EMDB—the Electron Microscopy Data Bank},
  journal      = {Nucleic Acids Research},
  year         = {2024},
  volume       = {52},
  number       = {D1},
  pages        = {D456--D465},
  doi          = {10.1093/nar/gkad1019}
}

@article{SingerYang2024,
    title={Alignment of density maps in {Wasserstein} distance},
    volume={4},
    doi={10.1017/S2633903X24000059},
    journal={Biological Imaging},
    author={Singer, Amit and Yang, Ruiyi},
    year={2024},
    pages={e5}
}

@inproceedings{LagerMordantMoscovich2026,
      title={Quantifying the noise sensitivity of the {Wasserstein} metric for images}, 
      author={Erik Lager and Gilles Mordant and Amit Moscovich},
      year={2026},
      booktitle = {Proceedings of the 43rd International Conference on Machine Learning (ICML)}
}

@inproceedings{Cuturi2013,
    author = {Cuturi, Marco},
    title = {Sinkhorn distances: lightspeed computation of optimal transport},
    year = {2013},
    publisher = {Curran Associates Inc.},
    address = {Red Hook, NY, USA},
    booktitle = {Proceedings of the 27th International Conference on Neural Information Processing Systems - Volume 2},
    pages = {2292--2300}, numpages = {9},
    location = {Lake Tahoe, Nevada},
    series = {NIPS'13}
}

@article{Bonneel2014,
    title={Sliced and Radon Wasserstein Barycenters of Measures},
    volume={51},
    ISSN={1573-7683},
    DOI={10.1007/s10851-014-0506-3},
    number={1},
    journal={Journal of Mathematical Imaging and Vision},
    publisher={Springer Science and Business Media LLC},
    author={Bonneel, Nicolas and Rabin, Julien and Peyré, Gabriel and Pfister, Hanspeter},
    year={2014},
    month=Apr,
    pages={22–45}
}

@inproceedings{Kolouri2019,
    author = {Kolouri, Soheil and Nadjahi, Kimia and Simsekli, Umut and Badeau, Roland and Rohde, Gustavo},
    booktitle = {Advances in Neural Information Processing Systems},
    editor = {H. Wallach and H. Larochelle and A. Beygelzimer and F. d\textquotesingle Alch\'{e}-Buc and E. Fox and R. Garnett},
    publisher = {Curran Associates, Inc.},
    title = {Generalized Sliced Wasserstein Distances},
    volume = {32},
    year = {2019}
}

@article{ShiSinger2022,
    title={Ab-initio contrast estimation and denoising of cryo-EM images},
    volume={224},
    ISSN={0169-2607},
    DOI={10.1016/j.cmpb.2022.107018},
    journal={Computer Methods and Programs in Biomedicine}, publisher={Elsevier BV},
    author={Shi, Yunpeng and Singer, Amit},
    year={2022},
    month=Sep,
    pages={107018}
}

\end{document}